\documentstyle[eqsecnum,aps]{revtex}
\begin{document}
\title{Bremsstrahlung radiation by a tunneling particle
}
\author{C.A. Bertulani$^{(a)}$, D. T. de Paula$^{(a)}$, 
and V.G. Zelevinsky$^{(b)}$ \ \footnote 
{Author e-mails: bertu@if.ufrj.br, dani@if.ufrj.br, zelevinsky@nscl.msu.edu}
}
\address{
$^{(a)}$Instituto de F\'{\i}sica, Universidade Federal 
do Rio de Janeiro,
21945-970 Rio de Janeiro, RJ, Brazil \\
$(b)$ Department of Physics and Astronomy
and National Superconducting Cyclotron Laboratory\\
Michigan State University, 
East Lansing, MI 48824-1321, USA 
}
\date{\today}
\maketitle

\begin{abstract}
We study the bremsstrahlung radiation of a tunneling charged particle in a
time-dependent picture. In particular, we treat
the case of bremsstrahlung during alpha-decay, which has been suggested as a
promissing tool to investigate the problem of tunneling times.
We show deviations of the numerical results from the semiclassical estimates.
A standard assumption of a preformed particle inside the well leads to
sharp high-frequency lines in the bremsstrahlung emission. These
lines correspond to ``quantum beats" of the internal part of
the wavefunction during tunneling arising from the interference of the
neighboring resonances in the well.
\end{abstract}

\vskip 0.5cm

Recent experiments \cite{Ka97} have triggered a great interest in the
phenomena of bremsstrahlung during tunneling processes which was discussed 
from different theoretical viewpoints in \cite{DG96,PB98,Ta98}. 
This can shed light on basic and still controversial 
quantum-mechanical problems of tunneling times
\cite{HS89}, especially in a complex and nonstationary environment.
It seems that $\alpha $-decay offers a unique possibility to study 
these fundamental questions. In ref. \cite{Ka97} it was claimed that the
bremsstrahlung spectrum in alpha-decay of $^{210}$Po could provide
information about the tunneling time. Their claim is based on
the comparison of the experimental spectra with a semiclassical calculation
of Dyakonov and Gornyi \cite{DG96}, which shows an interference pattern
arising from the contributions to bremsstrahlung from the inner, under the
barrier, and outer parts of the wave function. Papenbrock and Bertsch \cite
{PB98} performed a quantum-mechanical calculation for the bremsstrahlung in $%
\alpha $-decay in perturbation theory. They have shown that the contribution
from the tunneling wave function under the barrier is small, as well as the
appearance of interference effects in the spectrum. The greater effect, as
expected, arises from the Coulomb acceleration of the $\alpha $-particle
outside the nuclear well. Later, Takigawa et al. \cite{Ta98} compared the
quantum-mechanical calculations with classical and semiclassical results.
They conclude that subtle interferences of contributions 
from different parts of the wave function do indeed arise. However, in both
cases, the authors infer that the experimental data are not conclusive.
More exclusive experiments, with better statistics, are needed to
give a clearer understanding of the phenomena.

The theoretical approaches of refs. \cite{PB98,Ta98} assume a 
standard stationary
description of quantum tunneling which is very successful for
$\alpha $-decay lifetime and probabilities. However, a
time dependent picture of a decay process \cite{Fu84,Se94} may be
essential in understanding physics of the bremsstrahlung and similar processes,
in particular in
obtaining the appropriate bremsstrahlung spectrum. This can be
shown, for example, by looking into a case where there is no Coulomb
acceleration after the tunneling. In this letter we model the time evolution
of the wave function during the tunneling using
again $\alpha $-decay as an example. We do not attempt to compare
our results with the experimental data. The reason is simple: the $\alpha $%
-decay time, e.g., the lifetime of $^{210}$Po, is many orders of magnitude
larger than typical times for an $\alpha $-particle to traverse across the
nucleus. This implies that a stable numerical solution of the
Schr\"{o}dinger equation, keeping track simultaneously of 
fast oscillations in the well (``escape attempts") and
extremely slow tunneling, is virtually impossible: for $^{210}$Po it would
require about $10^{30}$ time steps in the iteration process. Instead, we
study the bremsstrahlung in high energy $\alpha $-decays, for which the
decay time is treatable numerically. This allows us to pay attention to
qualitatively new aspects of the bremsstrahlung in decay processes, and
compare with traditional aproaches.

For an $\alpha $-particle being accelerated from the turning point
to infinity, the classical bremsstrahlung can be calculated analytically.
Using the well known equations, see for example \cite{Ja75}, 
and integrating along the outward
branch of the Rutherford trajectory in a head-on collision we get for the
energy emitted by bremsstrahlung per frequency interval $d\omega $, in the
long-wavelength approximation,

\begin{equation}
dE\left( \omega \right) =\frac{8\pi \omega ^2}{3m^2c^3}Z_{eff}^2e^2\left|
p_r\left( \omega \right) \right| ^2\;.  \label{ene}
\end{equation}
where

\begin{equation}
p_r\left( \omega \right) =\frac{m a}{2\pi }e^{-\pi \nu /2}K_{i\nu }^{\prime
}\left( \nu \right) \;               \label{pclass}
\end{equation}
is the Fourier transform of the particle momentum,\ $m=m_N\cdot 4A/(A+4)$
is the reduced mass, and $Z_{eff}=\left( 2A-4Z\right) /(A+4)$ is the
effective charge of the alpha-particle + the daughter nucleus $(A,Z)$ in
the dipole approximation. In eq. (\ref{ene}), $p_r={\bf p\cdot }\widehat{%
{\bf r}}$, $a=Ze^2/E_\alpha $, $\nu =E_\alpha a/\hbar \mathrm{v}_0$, $%
E_\alpha $ is the energy of the alpha-particle, and $\mathrm{v}_0$ its
asymptotic velocity. The functions $K_{i\nu }\left( x\right) $ are the
modified Bessel functions of imaginary order, and $K_{i\nu }^{\prime }\left(
x\right) $ are their derivatives with respect to the argument. Dividing this
equation by the photon energy $E_\gamma =\hbar \omega $, we get the
differential probability per unit energy for the bremsstrahlung, i.e., $%
dP/dE_\gamma =\left( 1/E_\gamma \right) dE\left( \omega \right) /dE_\gamma $.

At low photon energies, we use the relation $\nu ^2e^{-\pi \nu }\left[
K_{i\nu }^{\prime }\left( \nu \right) \right] ^2\longrightarrow 1$ to show
that
\begin{equation}
\frac{dP}{dE_\gamma }=\frac 2{3\pi }\frac 1{E_\gamma }Z_{eff}^2\alpha \left( 
\frac{\mathrm{v}_0}c\right) ^2\;.  \label{prob}
\end{equation}

We solve the time dependent Schr\"{o}dinger
equation for the alpha-particle in a potential well + a Coulomb barrier, and
calculate the radial momentum from
\begin{equation}
p_r\left( t\right) =\frac \hbar i\int dr\;u^{*}(r,t)\frac{\partial u(r,t)}{%
\partial r}\;,  \label{pave}
\end{equation}
where $u(r,t)$ is the radial part of the total wavefunction.
A similar numerical study of the time evolution for the problem of the
$\alpha$-decay was performed by Serot {\sl et al.} \cite{Se94}.

The time-dependent Schr\"{o}dinger equation is solved, starting with an
$s$-wave wavefunction for an alpha particle confined in a radial well $V\left(
r\right) =-V_0$ for $r<R_0$, and $V\left( r\right) =-V_0+2Ze^2/R_0$ for $%
r\geq R_0$. We set $R_0=8.75$ fm, realistic for the
$\alpha+^{210}$Po case \cite{PB98}. At $t=0$ we switch the potential to a
spherical square well with depth $-V_0$ for $r<R_0$ and a Coulomb potential $%
2Ze^2/r$ for $r\geq R_0$ which triggers the tunneling.
At a given time $t$ the wave function is found by using the
standard method of inversion of a tri-diagonal matrix at each time step
increment $\Delta t$ (see, e.g. ref. \cite{Koo90}).The space is discretized
in steps of $\Delta x=0.05$ fm, and the time step used is $\Delta
t=m c\left( \Delta x\right) ^2/\hbar $. In Fig. 1 we plot the average
momentum for $\alpha $-energies of 23.1 MeV (1a) and 20 MeV
(1b). The $\alpha $-energies are changed by keeping the number of
nodes constant (in this case, 7 nodes), and varying the depth of the
potential from 14 to 18 MeV, respectively. The barrier height is 27.6 MeV.
The dotted line is the classical momentum for an $\alpha $-particle running
away from the closest approach distance. The dashed line is the average
momentum calculated according to eq. (\ref{pave}), but using only the part
of the wave function outside the barrier, i.e. the integral in (\ref{pave})
extending from $R_{cl}=$ $2Ze^2/E_\alpha $ to infinity. In this case, the
wave function entering (\ref{pave}) is normalized within this space 
range. 

The resulting ``quantum-mechanical'' momentum at the later stage is closer 
to the classical one. However, the quantum-mechanical momentum
increases slower than the classical one, due to the extended nature of the
particle's wave function. As a consequence, one expects that the Fourier
transform of the quantum momentum has its higher frequencies suppressed, as
compared to the classical case. The solid line represents the momentum
calculated according to (\ref{pave}), but including the full space range of
the wavefunction. One observes a wiggling pattern associated with the
interference of neighboring quasistationary states of the particle inside
the nucleus. 
The leaking of the inner part of the wave function creates an effective
oscillating dipole, which emits radiation. The same leaking creates the
perturbation mixing different resonance states.
The Fourier transform of the
particle momentum should therefore contain appreciable amplitudes
associated with this motion, as 
observed in Fig. 2, solid line. 
Asymptotically, the momenta calculated in all
different ways coincide at large times.

In Fig. 2 the dotted lines show the classical bremsstrahlung emission
probability, the dashed lines show the ``quantum-mechanical'' momentum
using only the part of the wave function outside the barrier, and the solid
lines are for the full space range of the wave function, as a function
of the photon momentum. The upper part of the graph is for $E_\alpha =23.1$
MeV, while the lower part is for $E_\alpha =20$ MeV. Since the 
quantum-mechanical 
momentum of the particle increases slower than the classical one,
the spectrum at larger photon energies is suppressed in comparison with the
classical one. Also, when one includes the whole wave function, the
bremsstrahlung spectrum is even more suppressed at large photon energies.
However, at very large photon energies ($E_\gamma \simeq 8.5$ MeV) the
spectrum shows a peak, revealing the interference between the different
components of the wave function in the well. The width 
$\Delta E$ of the peak is related
to the lifetime of the quasistationary state. For $E_{\alpha}=23$ MeV we find
$\Delta E=0.2$ MeV while for $E_{\alpha}=20$ MeV the width is narrow, $\Delta
E=0.02$ MeV. These values agree with the conventional Gamow formula for the 
width of the quasistationary state of the alpha-particle with the same energy 
and potential parameters.

In order to assess the relevance of the bremsstrahlung during tunneling 
and to shed light on the nature of the peaks in Fig. 2 we
wil now consider an $\alpha $-particle confined within a three-dimensional
square well with $V=-V_0$ for $r<R_0$, $V=U_0$ for
$R_0<r<R_1$, and $V=0$, otherwise.
We use the parameters $R_0=8.75$ fm, $R_1-R_0=1$ fm, $V_0=14$ MeV,
$U_0=27.6$ MeV, and $E_\alpha =24.2$ MeV.  For this 
problem a much simpler time-dependent solution can be
found. The time-dependent wave function is obtained from the expansion

\begin{equation}
u (r,t)=\int dE\,a(E)\;e^{iEt/\hbar }\; u_E(r)\;,  \label{tunbar}
\end{equation}
where $u_E(r)$ is the continuum (radial) 
wavefunction with energy $E$, normalized
to $4\pi \int dr u_E^{*}(r) u_{E^{\prime }}(r)=\delta \left(
E-E^{\prime }\right) $ and

\begin{equation}
a(E)=4\pi \int dr u_0(r)\; u_E(r)\;,  \label{tunbar1}
\end{equation}
where $u_0(r)$ is the radial wavefunction of the initial state. For a
square well plus barrier $u_0(r),$ $u_E(r),$ and $a\left( E\right) $
are given analytically. The time-dependent wavefunction is obtained from
eq. (\ref{tunbar}) by a simple integration.

In Fig. 3(a) we plot the momentum of the particle for this system as a
function of time. We observe a similar pattern as in Fig. 1(a), solid
curve, but with stronger oscillations, due to the quantum beats. In
Fig. 3(b) we show the corresponding bremsstrahlung spectrum. The resulting
pattern is very similar to that displayed in Fig. 2. It is important to
notice however that there is no Coulomb acceleration for this system. The
lower part of the energy spectrum is solely due to the tunneling through the
barrier, while the peak at higher energies is again due to the
interference during the tunneling process. 

We have compared the lower part of the spectrum in Fig. 3(b) 
with the result of
Dyakonov and Gornyi \cite{DG96}. They have obtained the bremsstrahlung
spectrum by a tunneling charge using perturbation theory and semiclassical
wave functions for the initial and final state of the particle. We have done
a similar calculation, but using the bound-state wave function $u_0(r)$
as the initial state. We obtained a bremsstrahlung spectrum for the soft
part of the spectrum which is by far smaller than that displayed in
Fig. 3(b). 
Moreover, the spectrum decays much faster than ours. This can be understood as
follows. The soft part
of the spectrum is due to the bremsstrahlung during tunneling. For a
particle in a one-dimensional tunneling motion through a square barrier,
Dyakonov and Gornyi's approach yields the spectrum given by
\begin{eqnarray}
\frac{dP}{dE_\gamma }=\frac 4{3\pi }\frac{U_0}{m c^2E_\gamma }%
Z_{eff}^2\alpha\  \exp( -2k_1d) \ \exp \left( -2\frac{E_\gamma d}{%
\hbar \mathrm{v}_1}\right) \left[ 1+\exp \left( -\frac{E_\gamma d}{\hbar 
\mathrm{v}_1}\right) \right] ^2,  \nonumber \\
\label{prob1}
\end{eqnarray}
where $U_0$ is the barrier height,  
$\mathrm{v}_1=\sqrt{2\left( U_0-E_\alpha \right) /m}$ is the imaginary
velocity of the
particle during tunneling, 
and $d$ is the barrier width.

This formula shows that the semiclassical bremsstrahlung spectrum of a
tunneling particle varies as $E_\gamma^{-1} $ for low photon energies and as $%
[\exp \left( -2E_\gamma d/\hbar \mathrm{v}_1\right)]/E_{\gamma}$ 
for high photon
energies. The slope parameter for high photon energies is given by $2d/\hbar 
\mathrm{v}_1$ as displayed by a dashed line in Fig. 3(b).
However, the spectrum obtained from the dynamical calculation, solid line
in Fig. 3(b), has a smaller slope parameter.
Besides, the spectrum shows pronounced peaks at large photon energies. Fig. 4,
where we show the scattering phase shift for the system ``well + barrier",
clarifies the origin of these peaks. The resonances at 15.7, 24.2 and 35.4 MeV
correspond to (bound or virtual) levels in the well. The amplitudes of 
eq. (\ref{tunbar1}) are presented in Fig. 4(b) where the resonance peaks are
also evident. The peaks at 9.1 MeV and 11.1 MeV, shown in Fig. 3(b), are due to 
interference of resonances at 15.7 MeV and 35.4 MeV with the initial state of
energy 24.2 MeV. The quantum beats apparently correspond to the energy 
differences ($24.2-15.7=8.5$) MeV and ($35.4-24.2=11.2$) MeV, respectively.
These values are close to the energies of the peaks appearing in Fig. 3(b).

The bremsstrahlung peaks associated with quantum beats are present in any
dynamical tunneling process, since an initial localized state always has some
overlap amplitude with neighboring states of the open well. The importance of
these peaks, or, equivalently, of admixture of neighboring resonances,
decreases if the initial state is a very sharp resonance as in the case of the
alpha-decay of $^{210}$Po studied in \cite{Ka97}. Until now it is poorly known
how an alpha-particle is preformed inside a nucleus. However, the initial wave
function must be of a localized nature, thus having a nonzero amplitude of
carrying a part of the wave function of an adjacent resonance. For high-lying
states, as shown above, this leads to the pronounced peaks in the
bremsstrahlung spectrum. The observation of those peaks would be valuable for
inferring the content of the initial wave function of a preformed
alpha-particle (or fission products).

The coupling to the radiation field can also influence the tunneling process.
Such effects require a fully quantum-mechanical approach which can be
formulated as follows. In the one-photon approximation for a
spinless particle, the Hamiltonian for the particle+photon system is given by

\begin{equation}
\widehat{H}=\widehat{H}_0\left( \widehat{{\bf r}},\widehat{%
{\bf p}}\right) -\frac{Z_{eff}e}{2m c}\;\widehat{{\bf p}}\cdot 
{\bf A}\left( \widehat{{\bf r}}\right) +\sum_{{\bf k},\lambda
}\hbar \omega _ka_{{\bf k},\lambda }^{\dagger }a_{{\bf k},\lambda }\;,
\label{hamilt}
\end{equation}
where $a_{{\bf k},\lambda }\;\left( a_{{\bf k},\lambda }^{\dagger
}\right) $ is the annihilation (creation) operator of photons with
momentum ${\bf k}$, and polarization $\lambda =1,2$. The middle term is the
particle-photon interaction and the last term is the photon Hamiltonian. The
electromagnetic field ${\bf A}$ is given by

\begin{equation}
{\bf A}({\bf r},t)=\sum_{{\bf k},\lambda }\sqrt{\frac{2\pi \hbar c^2%
}{\omega V}}\left[ a_{{\bf k},\lambda }\epsilon _{{\bf k},\lambda }e^{i%
{\bf k.r}-i\omega t}+a_{{\bf k},\lambda }^{\dagger }\epsilon _{{\bf %
k},\lambda }^{*}e^{-i{\bf k.r}+i\omega t}\right] \;,  \label{afield}
\end{equation}
where $V$ is the normalization volume, and $\epsilon _{{\bf k},\lambda }$
are the polarization vectors. In the mixed representation a system is given
by a function of \textbf{r} and the field is described in the occupation
number picture. Again, in the one-photon approximation the system
wave function can be written as

\begin{equation}
\left| \Psi \right\rangle =\Psi _0({\bf r},t)\left| 0\right\rangle +\sum_{%
{\bf k},\lambda }\Psi _{{\bf k},\lambda }^{\left( 1\right) }({\bf r}%
,t)\left| {\bf k},\lambda \right\rangle \;,  \label{onephot}
\end{equation}
where $\left| 0\right\rangle $ is the photon vacuum and $\left| {\bf k}%
,\lambda \right\rangle =$ $a_{{\bf k},\lambda }^{\dagger }\left|
0\right\rangle $ is the one-photon part. Inserting (\ref{hamilt}), (\ref
{afield}) and (\ref{onephot}) in the Schr\"{o}dinger equation we get the set
of coupled equations for  $\Psi _0({\bf r},t)$ and $\left\{ \Psi _{%
{\bf k},\lambda }^{\left( 1\right) }({\bf r},t)\right\} $

\begin{eqnarray}
i\hbar \frac{\partial \Psi _0}{\partial t} &=&\widehat{H}_0\Psi _0+\sum_{%
{\bf k},\lambda }g_{{\bf k},\lambda }\left( \widehat{{\bf p}}\cdot
\epsilon _{{\bf k},\lambda }\right) e^{i{\bf k.r}}\Psi _{{\bf k}%
,\lambda }^{\left( 1\right) }\;,  \nonumber \\
i\hbar \frac{\partial \Psi _{{\bf k},\lambda }^{\left( 1\right) }}{%
\partial t} &=&\left( \widehat{H}_0+\hbar \omega _k\right) \Psi _{{\bf k}%
,\lambda }^{\left( 1\right) }+\sum_{{\bf k},\lambda }g_{{\bf k}%
,\lambda }\left( \widehat{{\bf p}}\cdot \epsilon^* _{{\bf k},\lambda
}\right) \;e^{-i{\bf k.r}}\;\Psi _0\;.  \label{coupled}
\end{eqnarray}
where $g_{{\bf k},\lambda }$ incorporates all factors which depend on $%
\left( {\bf k},\lambda \right) $. The momentum operator $\widehat{{\bf %
p}}$ does not act on $e^{i{\bf k.r}}$, since ${\bf k\cdot \epsilon }=0$.

This approach goes back to the classical work by Pauli and Fierz \cite{PF38}.
It includes important feedback effects which renormalize the particle
trajectory by the coupling to the accompanying radiation field.
The coupled equations are to be solved for each photon momentum \textbf{k},
and polarization $\lambda $, with the initial conditions $\Psi _0({\bf r}%
,0)=\psi _0({\bf r})$ and $\Psi _{{\bf k},\lambda }^{\left( 1\right) }(%
{\bf r},0)=0$. By solving equations (\ref{coupled}) a full solution for
the particle wavefunction allows us to calculate the radiation spectrum.
The exact solution of these equations is crucial in the case when the photon
energy is comparable to the particle energy. For example, in the decay of
very low energy alphas this coupling may reduce the yield of high energy
photons. The reason is that when a particle emits a photon before tunneling,
it looses energy and this leads to a reduction of its barrier tunneling
probability. The exact results might be sensitive to the shape of the
barrier. We hope to come to this formulation of the problem elsewhere.

In conclusion, we have obtained the bremsstrahlung spectrum of a tunneling
particle (an $\alpha $-particle in a nucleus) by directly solving the
time-dependent Schr\"{o}dinger equation. As expected, we have found that
there are large deviations from the classical bremsstrahlung spectrum. We
have also seen that aproaches based on perturbation theory miss an important
piece of information, namely, the  time-dependent modification of the
particle wavefunction in the well during the decay time. This leads to
substantial emission of photons with frequencies close to those 
of quantum beats between neighboring resonances. 
This effect should be 
relevant in radiation emitted during $\alpha $-decay in nuclei. In a more
general case, the time dependence of the wave function of a tunneling
particle seems to deviate substantially from the spectrum calculated by
using perturbation theory with semiclassical wave functions. More experimental 
data on bremsstrahlung by a tunneling particle would be very welcome for
learning more about preformation states, dynamics of quantum beats, and
tunneling times. 
\bigskip

{\bf Acknowledgements}

This work was supported in part by the Brazilian funding agencies 
CNPq, FAPERJ, FUJB/UFRJ, and PRONEX, under contract 41.96.0886.00,
and by the NSF grant 96-05207.

\newpage
{\bf Figure Captions}

Fig. 1 - Average momentum (in MeV/$c$) for $\alpha $-energies of 23.1 MeV 
(a) and 20 MeV (b). In (a) the dotted line is the classical momentum for 
an $\alpha $-particle running
away from the closest approach distance. The dashed line is the average
momentum calculated according to eq. (\ref{pave}), but using only the part
of the wavefunction outside the barrier, i.e., the integral in (\ref{pave})
extending from $R_{cl}=$ $2Ze^2/E_\alpha $ to infinity. The solid line 
represents the momentum
calculated according to (\ref{pave}), but including the full space range of
the wavefunction. 

Fig. 2 - Classical bremsstrahlung emission
probability (dotted lines),  quantum-mechanical spectrum (dashed lines)
using only the part of the wave function outside the barrier, and using  the
full space wave function (solid
lines). The upper part of the graph is for $E_\alpha =23.1$
MeV, while the lower part is for $E_\alpha =20$ MeV. 

Fig. 3 - (a) Momentum (in MeV/$c$) of a particle in a square well plus barrier
as a
function of time. (b) Corresponding bremsstrahlung spectrum.

Fig 4 - (a) Phase-shifts ($l=0$) for the $\alpha + ^{210}$Po system, assuming a 
radial square-well of 8.75 fm size + a square barrier of 1 fm located at the
border
of the well. (b) Amplitudes for the overlap of the initial state ($E=24.2$
MeV) of the
closed well (barrier of infinitely large width) with neighboring states of
the open well.

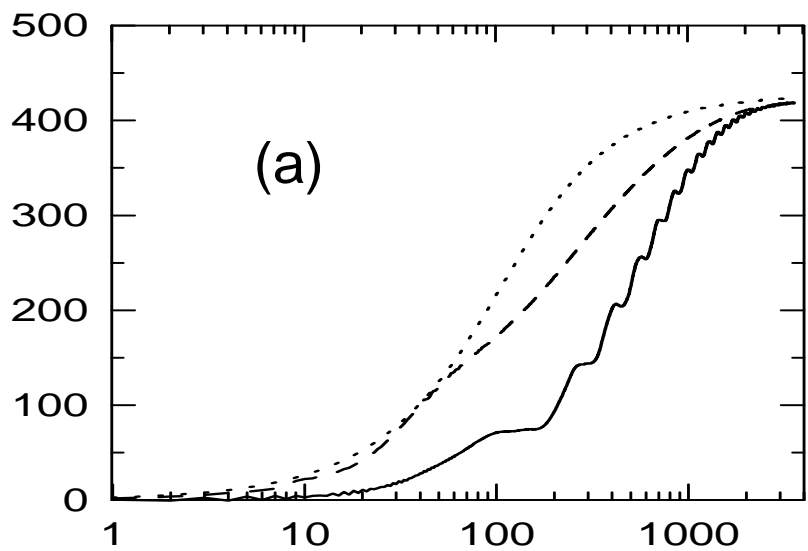
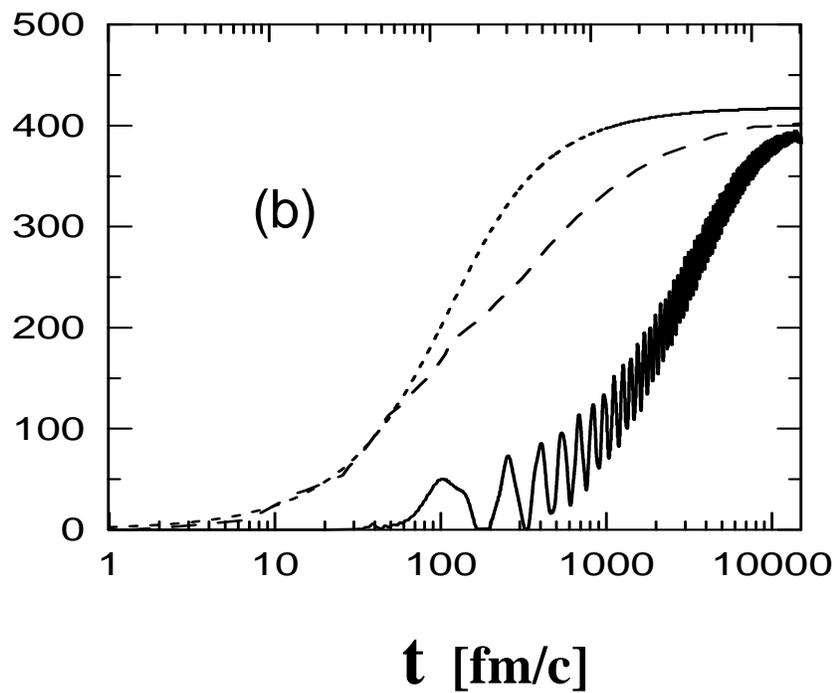
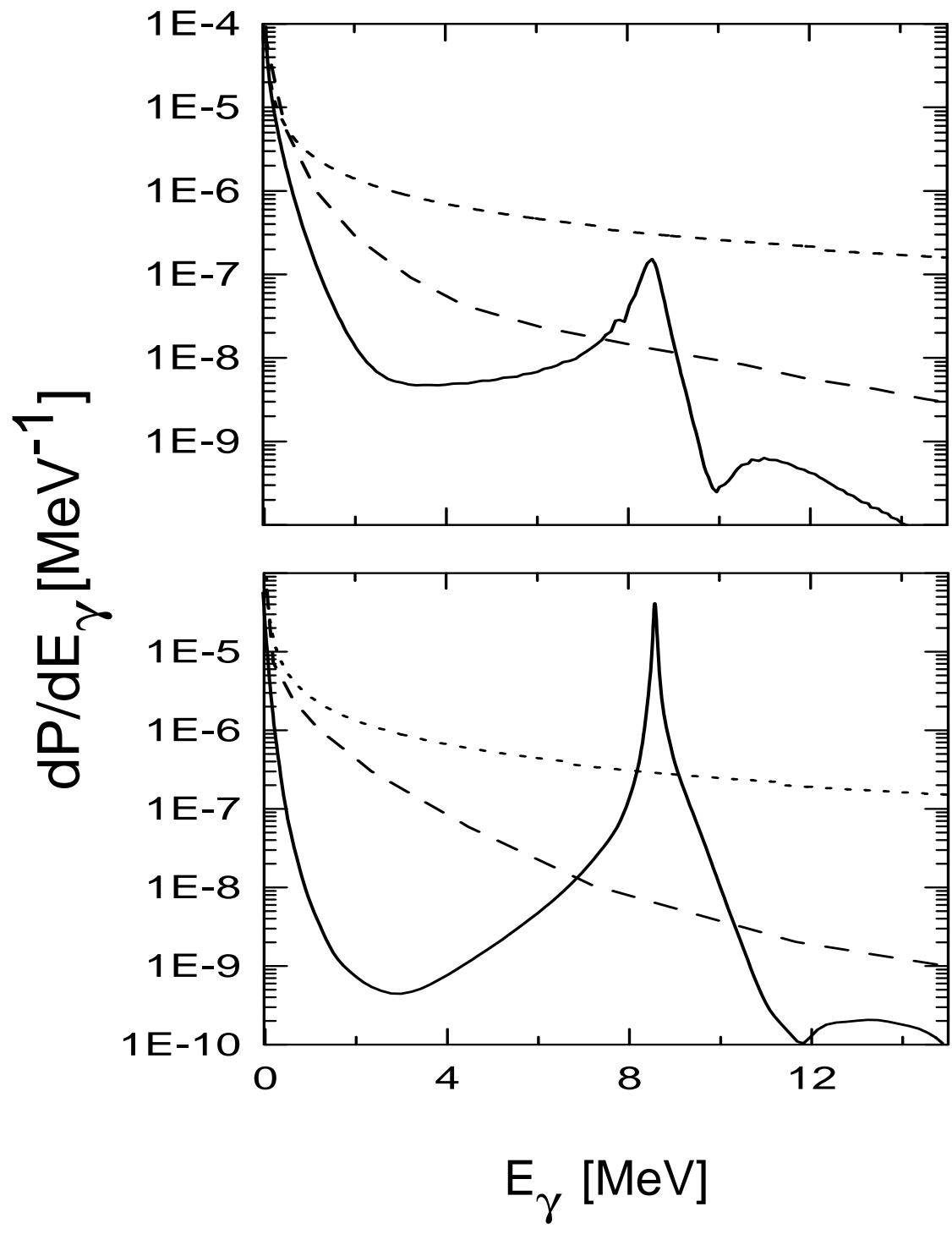
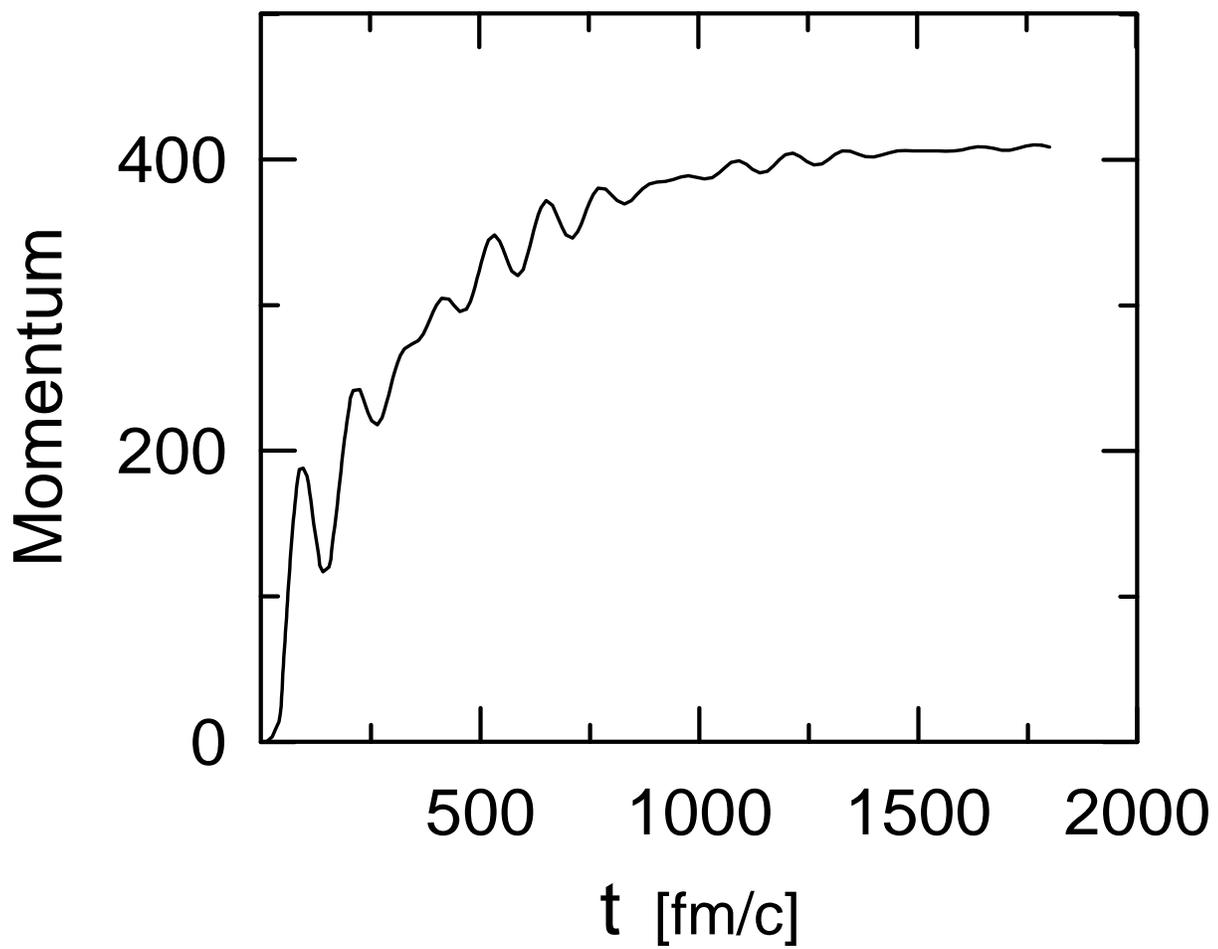
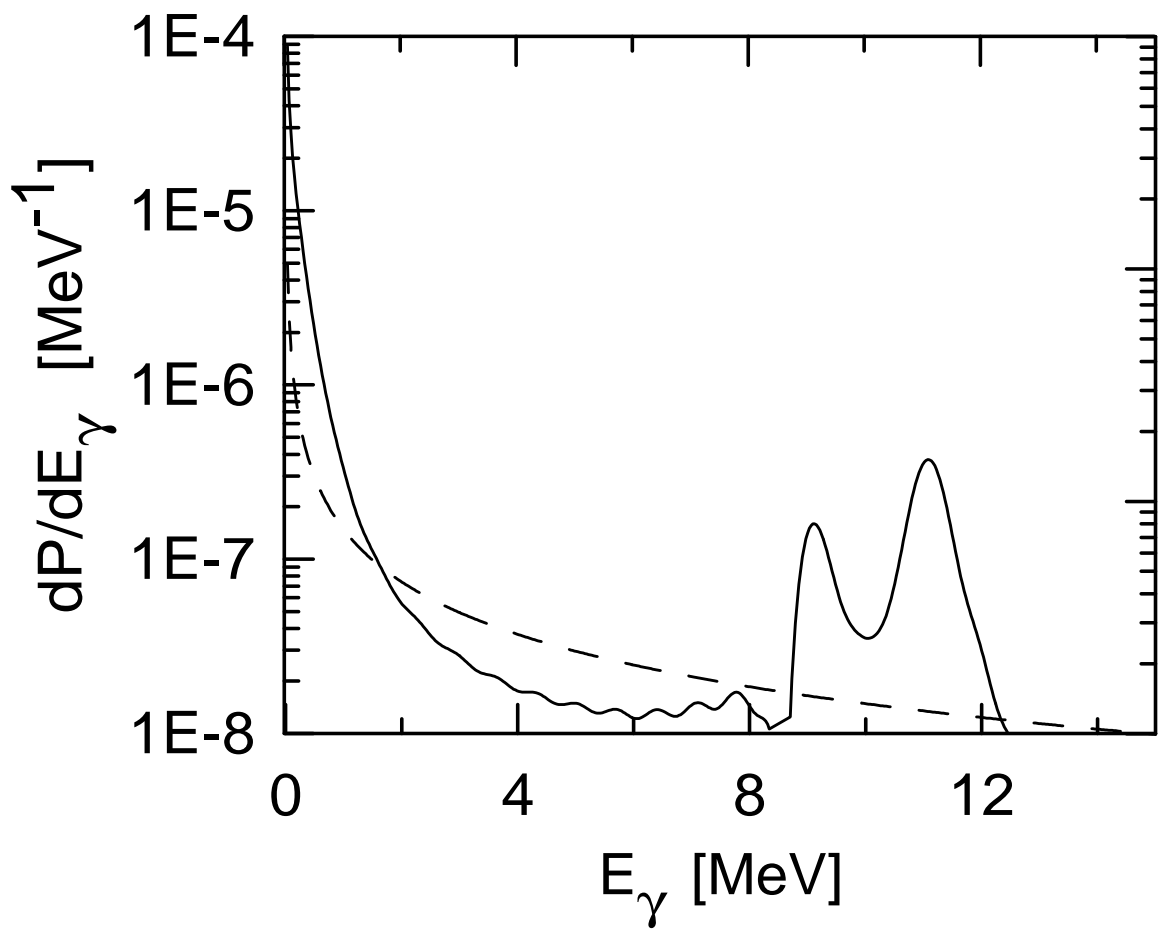
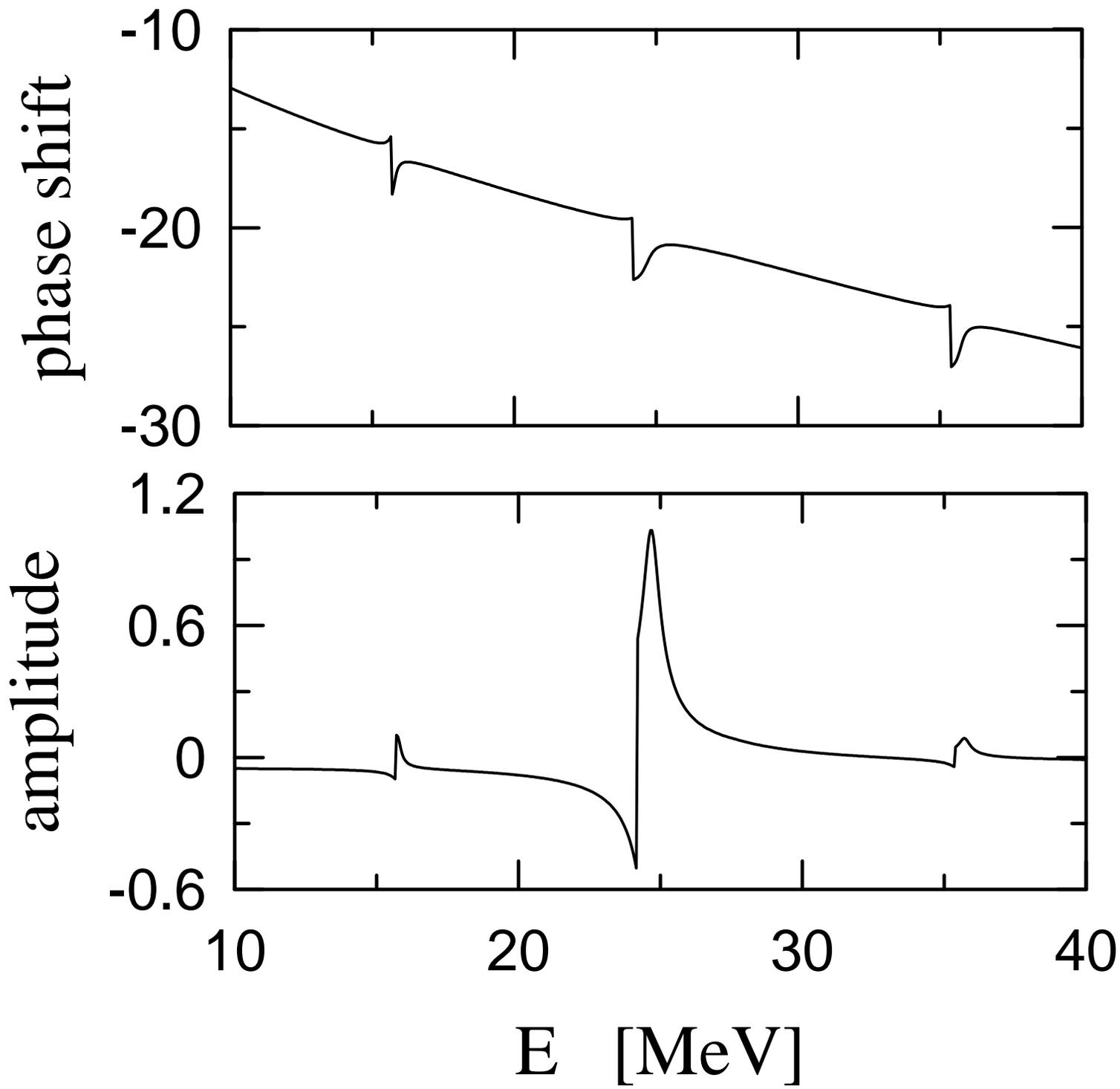
\end{document}